\documentclass[reprint,english,aip,bibnotes]{revtex4-1}
\usepackage[latin9]{inputenc}
\usepackage{amstext}
\usepackage{graphicx}
\usepackage{amssymb}
\usepackage{multirow}
\usepackage{dcolumn}
\usepackage{color}
\usepackage{amsmath}
\usepackage{babel}
\usepackage{hyperref}
\definecolor{BLUE}{rgb}{0,0,1}
\hypersetup{
     colorlinks   = true,
     citecolor    = BLUE,
     urlcolor = BLUE,
     linkcolor = black
}

\makeatletter
\newcommand {\aplt} {\ {\raise-.5ex\hbox{$\buildrel<\over\sim$}}\ }

\begin{document}

\title{Characterization and reduction of microfabrication-induced decoherence in superconducting quantum circuits}

\author{C. M. Quintana}
\affiliation{Department of Physics, University of California, Santa Barbara, California 93106, USA}
\author{A. Megrant}
\affiliation{Department of Physics, University of California, Santa Barbara, California 93106, USA}
\author{Z. Chen}
\affiliation{Department of Physics, University of California, Santa Barbara, California 93106, USA}
\author{A. Dunsworth}
\affiliation{Department of Physics, University of California, Santa Barbara, California 93106, USA}
\author{B. Chiaro}
\affiliation{Department of Physics, University of California, Santa Barbara, California 93106, USA}
\author{R. Barends}
\affiliation{Department of Physics, University of California, Santa Barbara, California 93106, USA}
\author{B. Campbell}
\affiliation{Department of Physics, University of California, Santa Barbara, California 93106, USA}
\author{Yu Chen}
\affiliation{Department of Physics, University of California, Santa Barbara, California 93106, USA}
\author{I.-C. Hoi}
\affiliation{Department of Physics, University of California, Santa Barbara, California 93106, USA}
\author{E. Jeffrey}
\affiliation{Department of Physics, University of California, Santa Barbara, California 93106, USA}
\author{J. Kelly}
\affiliation{Department of Physics, University of California, Santa Barbara, California 93106, USA}
\author{J. Y. Mutus}
\affiliation{Department of Physics, University of California, Santa Barbara, California 93106, USA}
\author{P. J. J. O'Malley}
\affiliation{Department of Physics, University of California, Santa Barbara, California 93106, USA}
\author{C. Neill}
\affiliation{Department of Physics, University of California, Santa Barbara, California 93106, USA}
\author{P. Roushan}
\affiliation{Department of Physics, University of California, Santa Barbara, California 93106, USA}
\author{D. Sank}
\affiliation{Department of Physics, University of California, Santa Barbara, California 93106, USA}
\author{A. Vainsencher}
\affiliation{Department of Physics, University of California, Santa Barbara, California 93106, USA}
\author{J. Wenner}
\affiliation{Department of Physics, University of California, Santa Barbara, California 93106, USA}
\author{T. C. White}
\affiliation{Department of Physics, University of California, Santa Barbara, California 93106, USA}
\author{A. N. Cleland}
\affiliation{Department of Physics, University of California, Santa Barbara, California 93106, USA}
\author{John M. Martinis}
\email{martinis@physics.ucsb.edu}
\affiliation{Department of Physics, University of California, Santa Barbara, California 93106, USA}

\date{\today}
\begin{abstract}
Many superconducting qubits are highly sensitive to dielectric loss, making the fabrication of coherent quantum circuits challenging. To elucidate this issue, we characterize the interfaces and surfaces of superconducting coplanar waveguide resonators and study the associated microwave loss. We show that contamination induced by traditional qubit lift-off processing is particularly detrimental to quality factors without proper substrate cleaning, while roughness plays at most a small role. Aggressive surface treatment is shown to damage the crystalline substrate and degrade resonator quality. We also introduce methods to characterize and remove ultra-thin resist residue, providing a way to quantify and minimize remnant sources of loss on device surfaces.
\end{abstract}
\maketitle

Improving the coherence times of superconducting qubits is of central importance for pushing quantum integrated circuits to a practical level of fault-tolerance for quantum computation,\cite{barends2014, chow2014,chow2012,fowler2012} as even moderate improvements to coherence can drastically reduce the overhead required for quantum error correction.\cite{fowler2012,barends2014} Substantial evidence has pointed to dielectric loss and fluctuations due to two-level system (TLS) tunneling defects\cite{phillips1972, anderson1972} as a source of energy relaxation in superconducting qubits and noise in sensitive superconducting photon detectors.\cite{martinis2005, gao2008, gao2008b, wang2009,barends2010,wisbey2010,vissers2010,paik2011,megrant2012, barends2013,chang2013,neill2013} These studies strongly suggest that TLS defects are located not in the bulk, but at the interfaces between device substrate, metal and vacuum, and that they can vary significantly with device materials, though the precise reason for this variation is usually a matter of speculation. The impact of TLS can be mitigated by increasing circuit dimensions,\cite{gao2008b,wenner2011,paik2011} but this strategy cannot be continued indefinitely. Nonetheless, the variable nature of the TLS-hosting circuit interfaces has not been carefully analyzed.

Here, we use superconducting resonators as sensitive probes to study TLS dielectric loss as a function of the processing used to construct these circuits, while concurrently analyzing the substrate-metal (S-M), substrate-vacuum (S-V), and metal-vacuum (M-V) interfaces. This allows us to separately extract the contributions of chemical contamination and induced disorder at the S-M interfaces of superconducting aluminum coplanar waveguide (CPW) resonators. In particular, we show how traditional processing methods can limit internal quality factors $Q_i$ to the range of $10^5 - 10^6$ at single-photon operating powers where TLS effects dominate. In addition, by characterizing resist residue we predict that without careful post-processing techniques, residual films of e-beam resist polymer on the vacuum interfaces may soon start to limit state-of-the-art superconducting qubit lifetimes.


We expect that measurements of resonator $Q_i$ will be predictive of dielectric loss in similarly fabricated superconducting qubits for two reasons: superconducting CPW resonator $Q_i$ are limited by energy relaxation and predict excitation lifetimes $T_1$ at single-photon powers,\cite{wang2008,wang2009,neill2013} and the large single-layer shunt capacitors of many superconducting qubits, such as the transmon, have interface participation ratios and hence dielectric losses comparable to those of a CPW.\cite{Koch2007,barends2013,steffen2010,birenbaum2014,wenner2011}

Transmon qubit capacitors have traditionally been fabricated using lift-off aluminum deposited together with their double-angle-evaporated Josephson junctions: first a ground plane is etched at the desired location of the qubit, and then electron-beam lithography is used to define the qubit pattern that is subsequently evaporated onto the etched substrate.\cite{frunzio2005,schreier2008,diCarlo2009} This means that the capacitor's S-M interface sees more processing than it would if the capacitor were formed by a subtractive etch alone. Improved coherence times have recently been found in transmons using lift-off metal for only a small fraction of the qubit,\cite{barends2013,chang2013} where the capacitors are first formed by an etch and the Josephson junctions are later evaporated after ion-milling the initial layer to remove native oxide and establish superconducting contact. In such a process, only a small fraction ($\sim1\,\%$) of the qubit self-capacitance is formed with lift-off  metal,\footnote{This neglects the Josephson junction itself, but experiments suggest that ultrasmall junctions exhibit remarkably low loss,\cite{paik2011,kim2011} presumably due to statistical avoidance of TLS.} and accordingly any extra dielectric loss induced by lift-off processing should be reduced by a similar factor. A systematic test to compare these two processes while keeping all other parameters constant, including metal type and growth conditions, has not been performed and could reveal information useful for improving qubit coherence.
\begin{figure*}[t!]
\centering
\includegraphics[width=1.69\columnwidth]{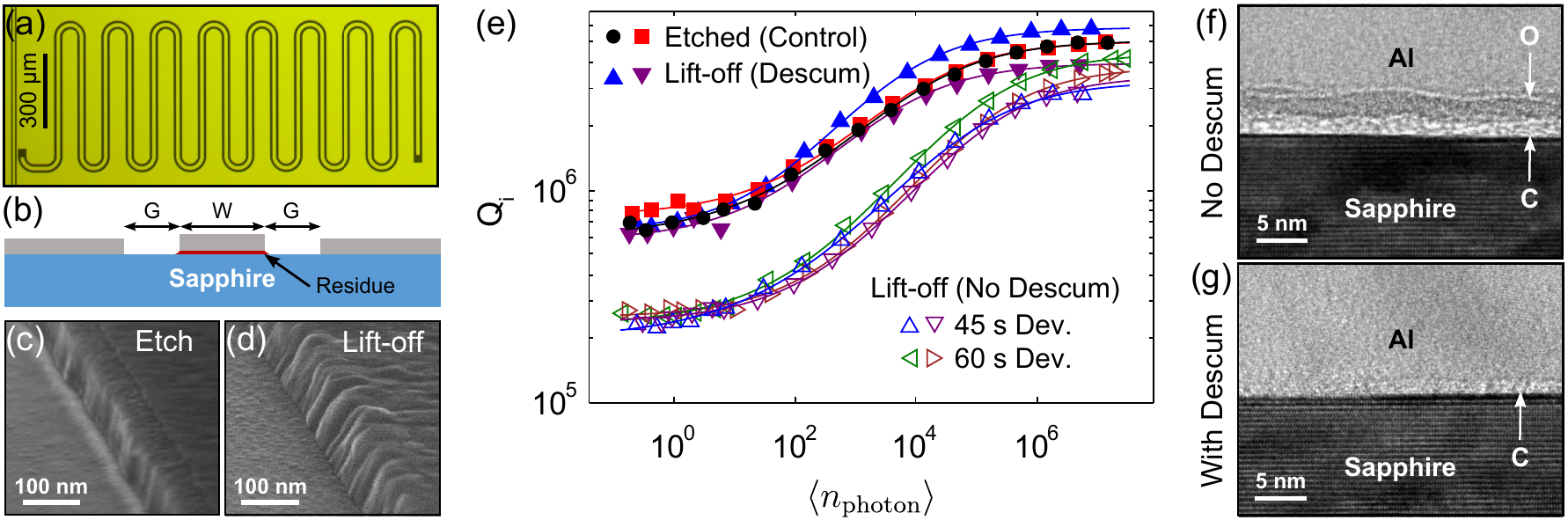}
\caption{\footnotesize Etch versus lift-off, and descum versus no descum: low-power $Q_i$ is degraded using lift-off without descum. (a) Optical micrograph of a ``hanging'' $\lambda/2$ CPW resonator capacitively coupled to a feedline (left). (b) Schematic of lift-off resonator cross-section. (c/d) SEM image of center trace edge of an etched/lift-off resonator. Etched metal sidewall is nearly vertical, with slight etching on the metal due to resist delamination, while lift-off metal sidewall has angle $\sim25^\circ$ from vertical. (e) Plot of internal quality factor $Q_i$ versus mean photon population for resonators made with lift-off with and without a pre-deposition descum, as well as for etched control resonators. Different marker types represent distinct resonators. Solid lines are guides to the eye. (f/g) Edge-on cross-sectional HRTEM image of S-M interface that saw processing similar to that of the lift-off resonators without/with descum. TEM sample ($\sim50\,\rm{nm}$ thick) prepared via gallium focused ion beam lift-out. Elemental peaks in C and O from qualitative EELS scans across the interfaces are indicated.
}
\end{figure*}

We perform such a controlled study by comparing the quality factors of CPW resonators fabricated with transmon-style lift-off versus a pure etch, both on the same chip, as follows. First, a polished $c$-plane sapphire wafer is solvent-cleaned, \footnote{5 min. sonication in acetone and then isopropanol [IPA]; spin dry.} and a base layer of aluminum is deposited in a high vacuum (HV) electron-beam evaporator after a gentle \emph{in situ} argon ion beam clean. Photolithography and a BCl$_3$/Cl$_2$  inductively coupled plasma (ICP) etch are then used to define $\lambda/2$ CPW resonators coupled to a feedline [Fig. 1(a)]. During this etch the ground plane slot of total width $W + 2G$ is defined for ``lift-off resonators''; the center traces of width $W$ for these resonators are deposited later with a lift-off technique [Fig. 1(b)]. During the ICP etch, as an on-chip control for the experiment the entire CPW structure (including the center trace) is defined for purely etched resonators. The center traces for the lift-off resonators are then defined in a way that mimics conventional transmon capacitor fabrication: the wafer is solvent-cleaned and then dehydration-baked\footnote{4 min. on a hotplate at 115\,$^\circ$C.}
and allowed to cool, after which a bilayer of 950K PMMA atop copolymer MMA(8.5)MAA e-beam resists are spun, each baked at $160\,^\circ\text{C}$ for 10 minutes. The center traces are then defined with e-beam lithography,\footnote{$100\,\text{kV}$, $2\,\text{nA}$ beam; dose $2000\,\mu\text{C}/\text{cm}^2$. Copolymer thickness $500\,\text{nm}$, PMMA thickness $300\,\text{nm}$.} after which the bilayer is developed with various development times in a 1:3 mixture of methyl isobutyl ketone (MIBK) to isopropanol (IPA), followed by a 10 second IPA dip and thorough nitrogen blow-dry. After development, the surface is optionally treated with a downstream oxygen ash descum before center trace deposition. During this descum,\footnote{3 min. in a Gasonics Aura 2000-LL downstream asher. We found it necessary to pre-clean the tool's chamber with a stronger recipe to obtain consistent descum results at the relatively low $150\,^\circ\text{C}$.} the substrate is heated to $150\,^\circ\text{C}$ and sees purely chemical cleaning with reactive oxygen, but not ions or plasma. We note that this cleaning is \emph{ex situ} with respect to the subsequent center trace deposition. The wafer is then transferred to and pumped overnight in the same HV e-beam evaporator used for the initial ground plane deposition, and the center traces are then deposited \emph{without} an \emph{in situ} clean. The metal is then lifted off in
N-Methyl-2-pyrrolidone (NMP) at $80\,^\circ\text{C}$ (3 hrs.) and cleaned in IPA.

The resonator chip is wirebonded into an aluminum sample box, which is mounted on the 50 mK stage of an adiabatic demagnetization refrigerator equipped with sufficient filtering and shielding so that radiation and magnetic vortex losses are negligible.\cite{Barends2011, megrant2012} All resonators had $W,G = 15,10\,\mu\text{m}$ with frequencies near $6\,\text{GHz}$. Using a feedline\cite{megrant2012} allows us to reproducibly extract $Q_i$ for multiple lift-off and etched resonators on the same chip. The resulting internal resonator quality factors are shown in Fig. 1(e). The decrease and saturation of $Q_i$ at low powers for all resonators is consistent with TLS-dominated loss. A clear difference (factor of 3) is observed in low-power $Q_i$ between the etched resonators and the lift-off resonators without descum. As seen in Fig. 1(e), the descum increases the low-power $Q_i$ back to or slightly below that of the control resonators. These measurements suggest that the edge profile of the resonators [Fig. 1(c/d)] had a negligible effect on loss at this level of coherence. It is also apparent that roughness of the S-M interface had a minimal effect on loss: the substrate under the center trace of the lift-off resonators was previously etched,\footnote{The aluminum ICP dry etch used to define the resonators etches $\sim 4\,\text{nm}$ into the sapphire substrate.} and is three times rougher than that under the center trace of the control resonators (0.3 versus 0.1 nm RMS roughness as measured by AFM).

To help understand the increased loss in the lift-off resonators, which we attribute to a contaminated S-M interface, we use cross-sectional high-resolution transmission electron microscopy (HRTEM) to examine the S-M interfaces of samples that saw similar\footnote{Differences arise due to fabricating multiple interfaces on a single TEM sample. The non-ashed S-M interface in the TEM sample saw processing temperatures up to $160\,^\circ\text{C}$, whereas the ashed S-M interface in the TEM sample and in the resonators only saw processing temperatures up to $115\,^\circ\text{C}$. As such, it is not certain if the upper AlO$_x$ sublayer was present in the non-ashed lift-off resonators. In addition, the ashed S-M interface in the TEM sample saw an initial coating of e-beam resist and subsequent strip before a second coating for e-beam lithography.} processing to the center traces of the lift-off resonators without/with the descum [Fig. 1(f/g)]. With no descum, we observe two sublayers at the S-M interface. Directly above the substrate is a film of average thickness 1.6 nm, presumably residual resist polymer, which shows a peak in carbon content when probed with electron energy loss spectroscopy (EELS). 
Above this, a $\sim2\,\text{nm}$ layer with similar phase contrast to aluminum oxide is observed, accompanied by a peak in oxygen content when probed with EELS. This layer is likely formed by a reaction of the unpassivated Al with resist and/or solvent residue either as the metal is evaporated onto the substrate, or during a later processing step when the wafer is heated. As such, it may contain contaminants such as hydrogen that may increase dielectric loss through the introduction of GHz-frequency TLS defects.\cite{martinis2005,khalil2013,holder2013,jameson2011} Oxide contamination from residue may be relevant to previous experiments finding that thermally oxidized submicron Josephson junctions are made significantly more stable by cleaning the substrate with oxygen plasma before metal deposition.\cite{koppinen2007,pop2012}

The S-M interface of the descummed substrate on the other hand shows a decreased average thickness\footnote{AFM scans of the substrate after e-beam resist exposure and development also reveal residual resist granules with widths of $\sim10 - 100\,\text{nm}$ and heights $\sim2 - 20\,\text{nm}$, even significantly above the e-beam clearing dose exposure, consistent with literature on PMMA.\cite{macintyre2009} However, we expect these granules to be negligible sources of dielectric loss compared with residual films, as the space they fill is negligible (areal fraction $\sim 0.5\,\%$ and equivalent uniform thickness $\aplt0.1\,\text{nm}$). Downstream ashing or UV-Ozone cleaning mostly removes these granules.} of carbon-containing residue and no observed peak in oxygen content. We note that our data is not sufficient to determine whether or not the decrease in carbon residue is in direct proportion to the decrease in resonator loss.

\emph{In situ} descum techniques such as ion beam cleaning may perform similarly to the downstream ash explored here. However, as this involves physical bombardment, a cleaning which is too aggressive might degrade the substrate quality at the interface. To test this hypothesis independently from questions of resist residue contamination, we fabricate etched superconducting $\lambda/4$ resonators whose substrates saw different strengths of \emph{in situ} argon ion beam cleaning prior to the base aluminum deposition: a weak clean (beam energy $200\,\text{eV}$, dose $\sim5\times10^{15}\,\text{cm}^{-2}$) and a stronger mill (beam energy $400\,\text{eV}$ and dose $\sim 5 \times 10^{17}\,\text{cm}^{-2}$). The stronger parameters are identical to those used to etch away native AlO$_x$ in the fabrication of Xmon transmon qubits\cite{barends2013} and similar to those used for substrate preparation in previous planar superconducting resonator experiments.\cite{geerlings2012} The resulting resonator quality factors are shown in Fig. 2(a), and display a power dependence consistent with TLS-dominated loss at low powers. We observe a clear difference (factor of 2) between the low-power internal quality factors, with the \emph{stronger} ion beam yielding a lower $Q_i$.
\begin{figure}[t!]
\begin{centering}
\includegraphics[width=1\linewidth]{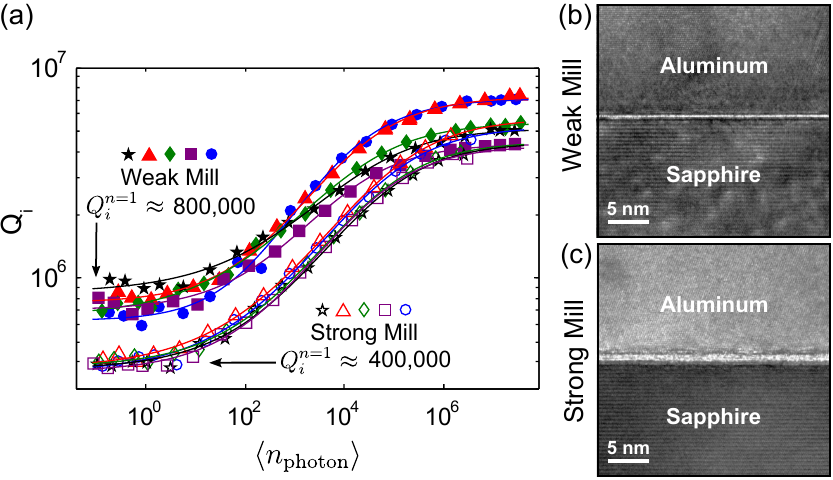}
\par\end{centering}
\caption{\footnotesize Comparison of weak and strong ion beam treatment, the latter inducing resonator degradation. (a) $Q_i$ of etched CPW resonators whose bare substrates saw weak or strong \emph{in situ} ion milling before base metal deposition (five resonators of each). (b/c) Cross-sectional HRTEM of S-M interface for weak/strong ion mill, showing thicker disordered interfacial layer for strong mill.}
\end{figure}

Fig. 2(b/c) shows cross-sectional HRTEM images of the S-M interface for the weak/strong ion beam treatments. The strong mill creates a $\sim 1.2\,\text{nm}$ interfacial layer, significantly thicker than the weakly-treated interface of unresolvable thickness. EELS reveals no measurable elemental peaks at either interface, including Ar, C, and O. We do not believe the uniform interface is an artifact of surface roughness, as AFM scans reveal no change in roughness between a bare and a strongly milled wafer, 
consistent with literature on sapphire.\cite{egert1992} We therefore attribute the excess loss to TLS induced by sapphire amorphization. Using finite-element COMSOL simulations,\cite{wenner2011} assuming a relative permittivity $\epsilon_r = 10$ for this layer we extract an intrinsic TLS loss tangent $\delta_\text{TLS}^0 \sim1\times10^{-2}$.\footnote{We only attribute participation to this layer at the S-M interface and not the partially etched S-V interface.}

Returning to Fig. 1(f), one may ask which of the interfacial sublayers of the lift-off resonators dominates the added loss, which could help inform future superconducting qubit fabrication. COMSOL simulations suggest that the added loss in the lift-off resonators without descum could be explained by a 2 nm thick interface with $\epsilon_r = 2$ (e.g., e-beam resist) and $\delta_\text{TLS}^0 = 3\times10^{-3}$, or with $\epsilon_r = 10$ (e.g., AlO$_x$) and $\delta_\text{TLS}^0 = 1.5\times10^{-2}$. In light of this, we more directly extract $\delta_\text{TLS}^0$ for the contamination by trapping it in a parallel plate capacitor, as illustrated in Fig. 3. The bottom plate of the capacitor (formed by part of the aluminum ground plane) is thoroughly cleaned, and then copolymer e-beam resist is spun, exposed and developed, after which the top capacitor plate is deposited along with the CPW center trace, trapping any residue. By forming a large (but physically small and thus lumped element) load capacitance $C_L$ at the end of a $\lambda/4$ CPW transmission line resonator, its net capacitance and loss tangent can be extracted from the shift in resonator frequency and the quality factor.
We derive analytical expressions for the load-dominated frequency shift and quality factor in the limit $C_L\gg C_\text{CPW}$, where $C_\text{CPW}$ is the total capacitance to ground of the CPW segment of the resonator. We model the lossy load capacitor as an ideal capacitor with an effective series resistance, $R_{ESR} = \tan\delta/\omega C_L$. For a single-dielectric $C_L$,
\begin{equation}
\omega_r - \omega_{\lambda/4} \approx \frac{2}{\pi C_L Z_r},~~~~ Q \approx \frac{1}{2\tan\delta}C_\text{CPW}C_\text{L}\omega^2 Z_r^2.
\end{equation}
Numerical SPICE simulations indicate that these expression are accurate to $1\%$ and $5\%$, respectively, for our experimental conditions. For a capacitor with more than one dielectric layer, voltage division allows one to extract the capacitance and $\tan\delta$ of one layer given those of the other. Using this fact and taking into account uncertainty in the thickness (3.3 - 4.0 nm),\footnote{Extracted by resistance measurements, TEM, and ellipsometry.} $\epsilon_r$ (9 - 10), and $\delta_\text{TLS}^0$ ($0.7-1.6\times10^{-3}$)\cite{martinis2005,pappas2011,khalil2013,deng2014}  of the bottom native\footnote{This oxide was originally thermally grown in a high-vacuum environment after ground plane deposition before exposure to atmosphere. A separate experiment in which residue was partially removed from the native oxide via a short descum lets us estimate an upper bound of $\sim 2.2\times 10^{-3}$ for $\delta_\text{TLS}^0$ of the native oxide at $\sim1\,\text{V/m}$, though it is possible\cite{khalil2013} that going to even lower powers would reveal a second, higher $\delta_\text{TLS}^0$.} AlO$_x$ layer, we conclude that the net $\delta_\text{TLS}^0$ of the contamination is in the range $1.6 - 3.6\times 10^{-3}$.
\begin{figure}[h!]
\begin{centering}
\includegraphics[width=.85\linewidth]{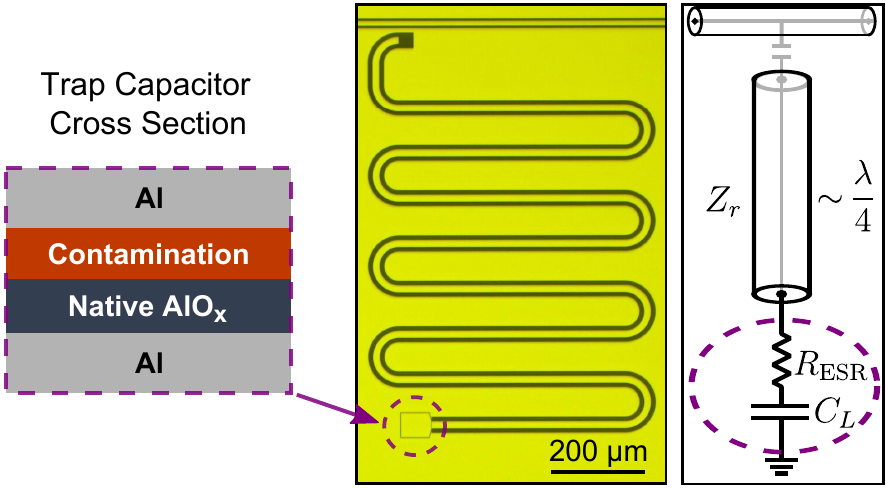}
\par\end{centering}
\caption{\footnotesize Schematic of ``trap capacitor'' experiment to characterize resist contamination. Left: capacitor cross-section. Middle: Optical micrograph of device. Right: Equivalent circuit diagram.
}
\end{figure}

We can compare this contamination loss tangent with that of bulk copolymer resist at low temperature and power. To measure $\delta_\text{TLS}^0$ of the bulk resist, we spin-coat CPW resonators with $500\,\text{nm}$ of copolymer, using the same $160\,^\circ\text{C}$ bake. From the resulting frequency shifts and low-power $Q_i$ of hanging $\lambda/4$ CPW resonators with multiple geometries, we extract (by simulating the capacitance per unit length of the coated resonator cross-sections in COMSOL) for the copolymer $\epsilon_r = 2.6\pm 0.1$ and $\delta_\text{TLS}^0 = (5.1\pm 0.3)\times 10^{-4}$. This loss is too small to quantitatively predict $\delta_\text{TLS}^0$ of the contamination layer extracted from the lift-off resonators or the trap capacitor. We conclude either that the lift-off loss comes from the oxide sublayer [Fig. 1(f)] or that $\delta_\text{TLS}^0$ of residue exposed to an e-beam is higher than that of the bulk polymer.
\begin{figure}[b]
\begin{centering}
\includegraphics[width=.79\linewidth]{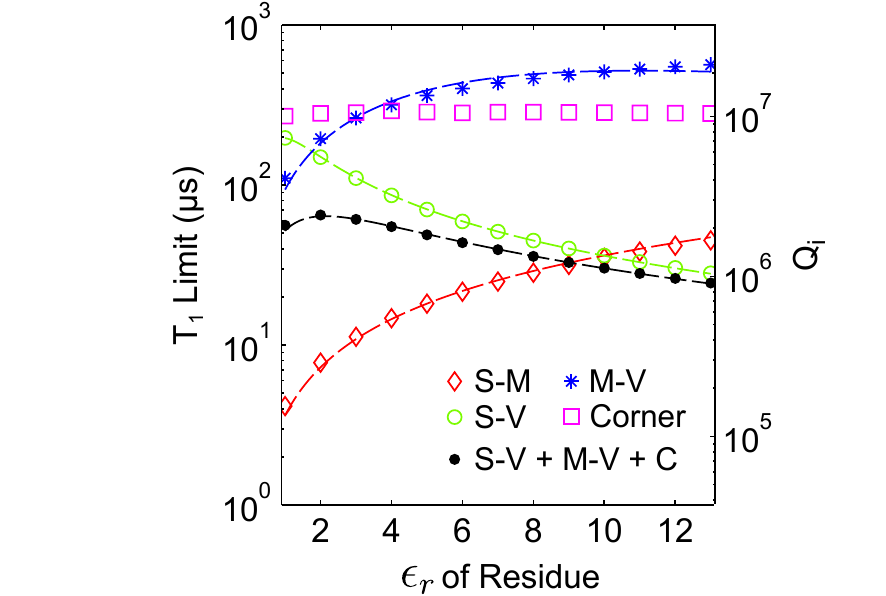}
\par\end{centering}
\caption{\footnotesize Markers: COMSOL simulations of the $T_1 = Q/\omega$ limit at $6\,\text{GHz}$ due to loss at the different types of CPW interfaces (considering separately ``corners'' where the interface types intersect\cite{wenner2011}), as a function of $\epsilon_r$ of the interface, assuming a $W,G=15,10\,\mu\text{m}$ untrenched CPW resonator with 3 nm thick interfaces of $\tan \delta = 2\times 10^{-3}$, with $\epsilon_r^\text{substrate}= 10$. Lines are fits to a simple model.\cite{wenner2011}
}
\end{figure}

To test the effect of exposure, we expose the resist-coated chip in deep ultraviolet (DUV) light sufficient to expose copolymer for development.\footnote{50 minutes in a DUV flood exposer, for a dose of $\sim 60\,\text{J}/\text{cm$^2$}$ with wavelengths between $200$ and $260 \,\text{nm}$}
We then repeat the measurements, observing a modestly increased bulk $\delta_\text{TLS}^0$ of $(7.7\pm 0.5)\times 10^{-4}$ with no measurable shift in $\epsilon_r$, still insufficient to quantitatively explain the contamination loss. Although PMMA has a very similar polymer fragmentation pathway and molecular weight distribution upon DUV exposure as it does for e-beam exposure,\cite{Uhl1998} ultra-thin polymer films may have significantly different properties from the bulk, due for example to interaction with the substrate.\cite{hartmann2002,sharp2003,leadley1997} We are thus unable to definitively conclude whether the polymer itself or contaminated AlO$_x$ dominated the loss in the lift-off resonators.

\begin{table}[t!]\footnotesize
\caption{\footnotesize Unexposed resist residue thicknesses of e-beam resist $\\$[MicroChem copolymer MMA(8.5)MAA) EL9] and i-line photoresist (Megaposit\texttrademark ~SPR955-CM) on sapphire, aluminum, and silicon, measured by ellipsometry (native oxides are accounted for).}
\begin{tabular}{l c c c}
\hline
\hline
Unexposed Resist Residue & Post-strip  & Post-descum \\
~~~~~~~($\pm 0.3$ nm)& (nm) &  (nm) &\\
\hline

\multirow{1}{*}{E-beam resist on Sapphire} & $4.2\footnote{5 min. ultrasonic agitation in acetone then IPA; spin dry.}$  & $0.0$\footnote{1 min. Gasonics downstream oxygen ash at $150\,^\circ\text{C}$.} \\
\multirow{1}{*}{E-beam resist on Sapphire} & $2.5\footnote{1 hour soak in heated NMP ($\sim70\,^\circ\text{C}$), followed by 5 min. ultrasonic agitation in heated NMP then IPA; spin dry.}$  & $0.0^\text{b}$ \\
\multirow{1}{*}{E-beam resist on Aluminum} & $2.8^\text{c}$  & $0.1^\text{b}$ \\

\multirow{1}{*}{E-beam resist on Silicon} & $3.9^\text{a}$   & $1.6/0.0$\footnote{UV-Ozone clean for 10/20 min., respectively.}\\
\multirow{1}{*}{E-beam resist on Silicon} & $0.4^\text{c}$   & $0.0^\text{b}$\\

  \multirow{1}{*}{Photoresist on Sapphire} & $0.2^\text{a}$   & $0.1^\text{b}$ \\
    \multirow{1}{*}{Photoresist on Silicon} & $0.1^\text{a}$   & $0.1^\text{b}$ \\

\multirow{1}{*}{Photoresist on Aluminum} & $0.6^\text{a}$   & $0.1^\text{b}$\\

\hline
\hline
\end{tabular}
\end{table}

The question of oxide contamination deserves further study, but in any case, it would be useful to detect and remove residual polymer, including on the vacuum interfaces. Previous interface participation simulations\cite{wenner2011} have focused on interfacial $\epsilon_r = 10$, for which the S-M and S-V interfaces participate equally and the M-V interface (i.e., surface oxide) is negligible. However, as shown in Fig. 4, the relative dielectric participation of the three CPW interface types depends strongly on the effective interfacial $\epsilon_r$. Note that S-M contamination is particularly detrimental at low $\epsilon_r$. We also see that post-processing residue on the substrate and even on the metal may start to limit coherence near the $100\,\mu\text{s}$ level for planar transmon qubits of modestly large size. It would therefore be useful to characterize the presence of residual films.

To detect and eliminate post-processing residue, we use variable angle spectroscopic ellipsometry to measure, on various surfaces, the ultra-thin residual films left by unprocessed e-beam resist and photoresist. Here, the resist is spin-coated onto a clean surface, baked, and then stripped. The results are summarized in Table I and are reproducible. We observe that the e-beam resist leaves substantially more residue than the photoresist, perhaps in part due to its higher bake temperature ($160\,^\circ\text{C}$ versus $95\,^\circ\text{C}$), justifying the assumption of $3\,\text{nm}$ for the S-V and M-V interfaces in Fig. 4, although the nature of leftover resist residue may depend strongly on any previous processing steps. Ultra-thin residue at the vacuum interfaces, then, may soon start to affect transmon lifetimes.

From Table I it is evident that some form of oxygen treatment is needed to completely remove the residual films. We note that UV-Ozone cleaning (row 4) is an effective alternative method that doesn't involve heating the substrate, which may be preferable for post-processing devices with Josephson junctions.\cite{pop2012} We do not observe any statistically significant change in $Q_i$ at the $\sim1\,\text{million}$ level after post-downstream-ashing the etched control resonators (which saw e-beam resist). Higher-quality epitaxial aluminum resonators\cite{megrant2012,megrant2014} would be necessary to detect improvement or degradation due to this vacuum-interface residue or post-downstream-ashing. We do however observe a significant decrease in low-power $Q_i$ (to $\sim200,000$) upon post-treating the etched resonators in a Technics PE-IIA oxygen plasma etch system ($300\,\text{mT O}_2$, $100\,\text{W}$ power for 30 sec., a common ``descum'' recipe), for reasons yet to be determined.

In conclusion, we have investigated the effects of interface processing on planar superconducting circuit coherence at the $Q_i = 10^5 - 10^6$ level at single-photon powers. At the S-M interface, we showed that contamination from resist residue and substrate damage from ion bombardment both significantly degrade resonator quality, while substrate roughness had a minimal effect. At the S-V and M-V interfaces, without oxygen treatment we observe post-processing residue at the vacuum interfaces that may start to limit planar superconducting qubit coherence at the level of $Q_i \sim$ several million, but find that post-treatment with aggressive oxygen plasma significantly degrades resonator quality. It would be worthwhile to test the effect of gentler types of post-ashing techniques on the coherence of superconducting qubits, about which there have only been anecdotal reports but no systematic study.\cite{slichter_thesis} It would also be important to investigate the influence, if any, of residual films and substrate damage on SQUID flux noise and other superconducting qubit dephasing mechanisms.\cite{sendelbach2008,choi2009,kumar2014,neill2013,omalley2014} Such post-processing studies will likely play an important role in further improving superconducting circuit coherence.

We thank B. Thibeault, U. Sharma, and C. Palmstr{\o}m for helpful discussions. Devices were fabricated at the UCSB Nanofabrication Facility, a part of the NSF-funded National Nanotechnology Infrastructure Network, and at the NanoStructures Cleanroom Facility. HRTEM and EELS were performed at the facilities of Evans Analytical Group in Sunnyvale, CA. This research was funded by the Office of the Director of National Intelligence (ODNI), Intelligence Advanced Research Projects Activity (IARPA), through Army Research Office Grant No. W911NF-09-1-0375. All statements of fact, opinion, or conclusions contained herein are those of the authors and should not be construed as representing the official views or policies of IARPA, the ODNI, or the U.S. Government.

\end{document}